\documentclass[pra, showpacs, twocolumn, floatfix,superscriptaddress]{revtex4}
\usepackage{graphicx}
\usepackage{epstopdf}
\usepackage{amsmath, amsfonts, amssymb, bm}

\def\mpik{Max-Planck-Institut f\"ur Kernphysik, Saupfercheckweg 1, D-69117
Heidelberg, Germany}

\def\ifa{Institute of Applied Physics, Academy of Sciences of Moldova,
Academiei str. 5, MD-2028 Chi\c{s}in\u{a}u, Moldova}
\begin{document}
\title{Cavity-output-field control via interference effects}
\author{Viorel \surname{Ciornea}}
\affiliation{\mpik}
\affiliation{\ifa}

\author{Mihai A. \surname{Macovei}}
\email{macovei@phys.asm.md}
\affiliation{\mpik}
\affiliation{\ifa}


\date{\today}
\begin{abstract}
We show how interference effects are responsible for manipulating
the output electromagnetic field of an optical micro-resonator 
in the good-cavity limit. The system of interest consists in a 
moderately strongly pumped two-level emitter embedded in the 
optical cavity. When an additional weaker laser of the same 
frequency is pumping the combined system through one of the 
resonator's mirror then the output cavity electromagnetic field 
can be almost completely suppressed or enhanced. This is due to the 
interference among the scattered light by the strongly pumped atom 
into the cavity mode and the incident weaker laser field. The result 
applies to photonic crystal environments as well.
\end{abstract}
\maketitle

\section{Introduction}
Present and future quantum technologies require various tools allowing for complete or partial control of 
the quantum mechanical interaction between light and matter. Therefore, quite significant 
amount of works are dedicated to this issue. Particularly, light interference is an widely 
investigated topic and, no doubts, its importance for various applications is enormous 
\cite{bw,gsa_book,al_eb,ficek,kmek}. Due to quantum interference effects for instance 
elimination of spectral lines or complete cancellation of the spontaneous decay can occur. 
Spatial interference shows interesting features as well \cite{gsa_book,ficek,kmek}. 
Furthermore, suppression of the resonance fluorescence in a lossless cavity was demonstrated in 
\cite{car1} whereas cavity-field-assisted atomic relaxation and suppression of resonance fluorescence
at high intensities was shown in \cite{sup2}. Inhibition of fluorescence in a squeezed vacuum was 
demonstrated in \cite{su3} while suppression of Bragg scattering by collective interference of spatially 
ordered atoms within a high-Q cavity mode was demonstrated, respectively, in \cite{sup3}. On the other 
side, cavity-enhanced single-atom spontaneous emission was observed in Ref.~\cite{har1} while 
suppression of spontaneous decay at optical frequencies was shown in \cite{har2}. The control of the 
spontaneous decay as well as of the resonance fluorescence is of particular interest for quantum 
computation processes \cite{chuang} where, in addition, highly correlated photons are required \cite{cbm}. 
Combining few coherent driving sources one can achieve a further degree of control of the atom's quantum 
dynamics. Actually, the bichromatic driving of single atoms was intensively investigated recently emphasizing 
interesting interference phenomena. In particular, the resonance fluorescence of a two-level atom in a strong 
bichromatic field was analyzed in \cite{fic} and the response of a two-level system to two strong fields was 
experimentally studied in \cite{mrt}, correspondingly. The decay of bichromatically driven atom in a 
cavity was investigated in \cite{wa}. Broadband high-resolution x-ray frequency combs were obtained via 
bichromatically pumping of three-level $\Lambda-$type atoms \cite{schk}. Moreover, bichromatic driving of a 
solid-state cavity quantum electrodynamics system was investigated in Ref.~\cite{vuc}. Finally, photonic 
crystal's influence on quantum dynamics of pumped few-level qubits was investigated in detail as well 
\cite{john,me_chk,gx}.

The above mentioned works may be of particular relevance in a quantum network \cite{kimble,rempe}, for instance. 
Related systems have been already proven to act as an optical diode \cite{joerg,ch} - an important 
ingredient in a quantum network. Since a precise control over system's properties is highly required in such a 
network, here, we investigate the feasibility of controlling the cavity output electromagnetic field in a system 
consisting of a moderately strongly pumped two-level emitter. If a second coherent driving is applied through 
one of the mirrors and perpendicular to the first laser-beam then the output cavity field can be almost 
completely inhibited in the good-cavity limit. Notice that the lasers are in resonance with the cavity mode 
frequency. We have found that the interference between the second weaker light beam and the light 
scattered by the two-level emitter into the cavity mode due to stronger pumping is responsible for the 
suppression effect. The destructive interference can be turned into constructive one (or viceversa) via 
varying the phase difference of the applied lasers. Furthermore, the inhibition requires the laser frequency 
to be out of atomic frequency resonance while for photonic crystals surroundings it can be even on resonance. 

The article is organized as follows. In Sec. II we describe the analytical approach and the system of interest, while in
Sec. III we analyze the obtained results. The summary is given in Sec. IV.

\section{Quantum dynamics of a pumped two-level atom inside a 
driven microcavity}
\begin{figure}[t]
\includegraphics[width = 8cm]{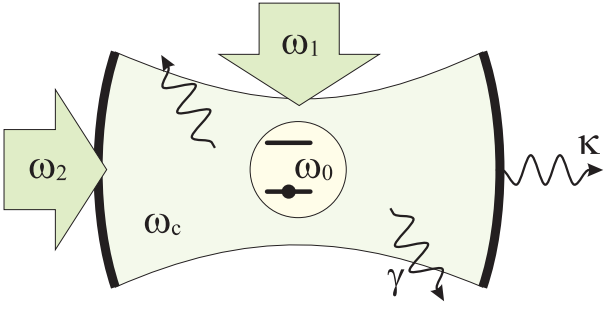}
\caption{\label{fig1} (color online) The schematic of the model: 
A two-level emitter possessing the transition frequency $\omega_{0}$ 
embedded in a single-mode ($\omega_{c}$) micro-cavity is pumped with 
an intense laser field of frequency $\omega_{1}$.
A second coherent source of frequency $\omega_{2}$ is driving the 
entire system through one of the mirrors. $\gamma$ is the single-atom 
spontaneous decay rate, while $\kappa$ describes the cavity photon leaking 
rate, respectively.}
\end{figure}

The Hamiltonian describing a two-level atomic system having the 
transition frequency $\omega_{0}$ and interacting with a strong coherent 
source of frequency $\omega_{1}$ while embedded in a   pumped micro-cavity 
of frequency $\omega_{c}$, in a frame 
rotating at $\omega=\omega_{1}=\omega_{2}$ 
(See Figure \ref{fig1}), is:
\begin{eqnarray}
H&=&\hbar\Delta S_{z} + \hbar\delta a^{\dagger}a 
+ \hbar g(a^{\dagger}S^{-}+aS^{+}) \nonumber \\
&+&\hbar\Omega(S^{+}e^{i\phi_{1}} + S^{-}e^{-i\phi_{1}})+ \hbar\epsilon(a^{\dagger}e^{i\phi_{2}} + ae^{-i\phi_{2}}), \nonumber \\
\label{HM}
\end{eqnarray}
where $\Delta=\omega_{0}-\omega$, and $\delta=\omega_{c}-\omega$. 
In the Hamiltonian (\ref{HM}) the components, in order of appearance, 
describe the atomic and the cavity free energies, the interaction 
of the two-level emitter with the micro-cavity mode and the atom's 
interaction with the first laser field with $\Omega$ being the 
corresponding Rabi frequency, and respectively, the interaction of 
the second driving field with the cavity mode with $\epsilon$ 
being proportional to the input laser field strength amplitude. 
The atomic bare-state operators $S^{+}=|e\rangle \langle g|$ and
$S^{-}=[S^{+}]^{\dagger}$ obey the commutation relations for su(2) 
algebra, i.e., $[S^{+},S^{-}]=2S_{z}$ and $[S_{z},S^{\pm}]=\pm S^{\pm}$. 
Here, $S_{z}=(|e\rangle \langle e|-|g\rangle \langle g|)/2$ is the 
bare-state inversion operator. $|e\rangle$ and $|g\rangle$ are, 
respectively, the excited and ground state of the atom while 
$a^{\dagger}$ and $a$ are the creation and the annihilation
operator of the electromagnetic field (EMF) in the resonator, and 
satisfy the standard bosonic commutation relations, namely, 
$[a,a^{\dagger}]=1$, and $[a,a]=[a^{\dagger},a^{\dagger}]=0$ \cite{zb_sc,ger}.
$\{\phi_{1},\phi_{2}\}$ are the corresponding phases of the coherent 
driving sources.

We shall describe our system using the laser-qubit semiclassical 
dressed-state formalism defined as \cite{ficek}: 
\begin{eqnarray}
|+\rangle&=&\sin\theta|g\rangle+\cos\theta|e\rangle,\nonumber\\
|-\rangle&=&\cos\theta|g\rangle-\sin\theta|e\rangle, \label{drs}
\end{eqnarray}
with $\tan 2\theta=2\Omega/ \Delta$. Applying this transformation to 
(\ref{HM}) one arrives then at the following dressed-state Hamiltonian
\begin{eqnarray}
H=H_0&+&\hbar g(\cos^2\theta R^--\sin^2\theta R^+)a^\dagger e^{-i\phi}
        \nonumber\\
     &+&\hbar g(\cos^2\theta R^+-\sin^2\theta R^-)a e^{i\phi},
        \label{HI}
\end{eqnarray}
with
\begin{eqnarray}
H_0=\hbar\bar\Omega R_{z}+\hbar\delta a^{\dagger} a
   +\hbar \epsilon(a^{\dagger} + a) + \hbar R_{z}(g^{\ast}_{0} a^\dagger + g_{0}a). \nonumber \\
 \label{HII}
\end{eqnarray}
Here, $\bar\Omega=\sqrt{\Omega^2+(\Delta/2)^2}$ while $g_{0}=(g/2)\sin{2\theta}e^{i\phi}$, $g^{\ast}_{0}=(g/2)\sin{2\theta}e^{-i\phi}$ and 
$\phi=\phi_{2}-\phi_{1}$. We employed also $S^{\pm}=\tilde S^{\pm}e^{\mp i\phi_{1}}$ and $a^{\dagger}=\tilde a^{\dagger}e^{-i\phi_{2}}$ with 
$a=[a^{\dagger}]^{\dagger}$ in the Hamiltonian (\ref{HM}) and dropped the tilde afterwards. The new quasi-spin 
operators, i.e., $R^{+}=|+\rangle\langle-|$, $R^{-}=[R^{+}]^{\dagger}$ 
and $R_{z}=|+\rangle\langle +| - |-\rangle\langle-|$ are operating in 
the dressed-state picture. They obey the following commutation relations: 
$[R^+,R^-]=R_z$ and $[R_z,R^\pm]=\pm 2R^\pm$. 

Considering that $\delta \ll \bar \Omega$ the last two terms in Eq.~(\ref{HI}) 
can be ignored under the secular approximation. Therefore, the master equation 
describing the laser-dressed two-level atom inside a leaking pumped resonator 
and damped via the vacuum modes of the surrounding EMF reservoir is:
\begin{eqnarray}
\frac{d}{dt}\rho(t) &+&\frac{i}{\hbar}[H_{0},\rho]
=- \kappa [a^{\dagger},a\rho] -\Gamma_{0}[R_{z},R_{z}\rho]
 \nonumber \\
&-& 
\Gamma_{+}[R^{+},R^{-}\rho] - \Gamma_{-}[R^{-},R^{+}\rho]  + H.c.. \label{ME}
\end{eqnarray}
Here 
\begin{eqnarray*}
\Gamma_{0} &=& (\gamma_{0}\sin^{2}{2\theta}+\gamma_{d}\cos^{2}{2\theta})/4, 
\nonumber \\
\Gamma_{+} &=& \gamma_{+}\cos^{4}{\theta} + (\gamma_{d}/4)\sin^{2}{2\theta},
\nonumber \\
\Gamma_{-} &=& \gamma_{-}\sin^{4}{\theta} + (\gamma_{d}/4)\sin^{2}{2\theta}. 
\end{eqnarray*}
Respectively, $\gamma_{0}=\pi\sum_{k}g^{2}_{k}\delta(\omega_{k}-\omega)$ and $\gamma_{\pm}=\pi\sum_{k}g^{2}_{k}\delta(\omega_{k}-\omega \mp 2\bar \Omega)$ 
are the single-atom spontaneous decay rates being dependent on the density of modes, $g_{k}$, at the dressed-state frequencies $\{\omega, \omega \pm 2\bar \Omega\}$, 
while $\gamma_{d}$ signifies the pure dephasing rate. In free space one has that $\gamma_{0}=\gamma_{\pm} \equiv \gamma$. Note that the master 
equation (\ref{ME}) was obtained either under the intense-field condition or far off detuned field, i.e., 
it is valid when $\bar \Omega \equiv \sqrt{\Omega^2+(\Delta/2)^2} \gg \{\delta,g,\epsilon,\Gamma_{0},\Gamma_{\pm}\}$.

The equations of motion for the variables of interest can be easily 
obtained from the Master Equation (\ref{ME}). Therefore, the quantum dynamics is described by the following system of linear differential equations:
\begin{eqnarray}
\frac{d}{dt}\langle a^{\dagger}a\rangle&=&ig_{0}\langle R_{z}a\rangle + i\epsilon\langle a\rangle
-ig^{\ast}_{0}\langle R_{z}a^\dagger\rangle
-i\epsilon\langle a^{\dagger}\rangle \nonumber \\
&-& 2\kappa\langle a^{\dagger} a\rangle,\nonumber\\
\frac{d}{dt}\langle R_{z}a\rangle&=&-(\kappa + i\delta + 2\Gamma_{+} + 2\Gamma_{-})\langle R_{z} a\rangle
    \nonumber\\
&-&2(\Gamma_{+} - \Gamma_{-})\langle a\rangle - i\epsilon\langle R_{z}\rangle -  ig^{\ast}_{0},
    \nonumber\\
\frac{d}{dt}\langle R_{z}a^{\dagger}\rangle&=&-(\kappa - i\delta + 2\Gamma_{+} + 2\Gamma_{-})
\langle R_{z}a^{\dagger}\rangle \nonumber\\
&-&2(\Gamma_{+} - \Gamma_{-})\langle a^{\dagger}\rangle + i\epsilon\langle R_{z}\rangle + ig_{0},
    \nonumber\\
\frac{d}{dt}\langle a \rangle&=&-(\kappa + i\delta)\langle a\rangle - ig^{\ast}_{0}\langle R_{z}\rangle - i\epsilon, \nonumber\\
\frac{d}{dt}\langle a^{\dagger}\rangle&=&-(\kappa - i\delta)\langle a^{\dagger}\rangle + ig_{0}\langle R_{z}\rangle + i\epsilon, \nonumber \\
\frac{d}{dt}\langle R_{z}\rangle &=& -2(\Gamma_{-}+ \Gamma_{+})\langle R_{z}\rangle + 2(\Gamma_{-} - \Gamma_{+}).
\label{sseq1}
\end{eqnarray}
In the system of equations (\ref {sseq1}), we have used the trivial condition $R^{2}_{z}=1$ 
which is the case for a single-qubit system.

In the following Section, we shall discuss our results, i.e., the possibility 
of inhibiting the cavity output field in the steady-state via interference effects.

\section{Output cavity field control}
\begin{figure}[t]
\includegraphics[width = 8cm]{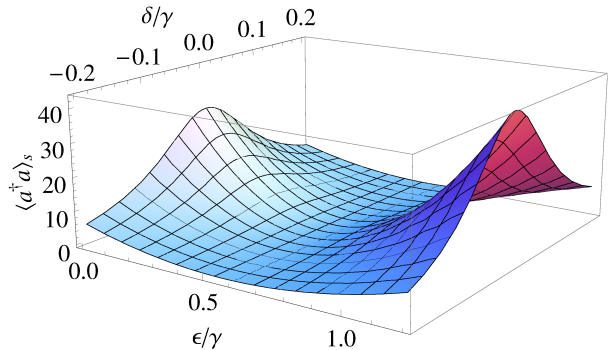}
\caption{\label{fig2} (color online) 
The steady-state dependence of the micro-cavity mean photon number 
$\langle a^\dagger a\rangle_s$ versus the variables $\epsilon/\gamma$ and 
$\delta/\gamma$. Other parameters are: $\gamma_d/\gamma = 0.01$, 
$\kappa/\gamma = 0.1$, $g/\gamma = 2$, $\Delta/\Omega = 3$ and $\phi=0$.}
\end{figure}

One of the solutions of system (\ref {sseq1}) in the steady-state represents the 
mean-photon number in the micro-cavity mode, namely:
\begin{eqnarray}
\langle a^\dagger a\rangle_s=A\epsilon^2+B\epsilon+C.\label{apm}
\end{eqnarray}
For $\delta=0$ and $\gamma_{0}=\gamma_{\pm} \equiv \gamma$, the coefficients 
$A$, $B$ and $C$ are given by the following expressions:
\begin{eqnarray}
A&=&\frac{1}{\kappa^2}, \nonumber \\
B&=&-\frac{2 g \gamma\Delta\Omega\cos{\phi}}
{\kappa^2(\gamma \Delta^2+2(\gamma+\gamma_d)\Omega^2)},\nonumber\\
C&=&\frac{g^2\Omega^2}{\kappa^2(\gamma\Delta^2+2(\gamma+\gamma_d)
\Omega^2)}\nonumber\\
&\times&\frac{\gamma(\kappa+2\gamma)\Delta^2+2\kappa(\gamma 
+\gamma_d)\Omega^2}{(\kappa+2\gamma)\Delta^2+4(\kappa+\gamma 
+\gamma_d)\Omega^2}. \label{kf}
\end{eqnarray}
Because of the quadratic dependence on $\epsilon$, the minimum value 
of the mean-photon number is:
\begin{eqnarray}
\langle a^{\dagger}a\rangle_{s}^{\min}=C-\frac{B^2}{4A}.
\label{min}
\end{eqnarray}
The above value is achieved at: 
$$\epsilon^{\min}=-B/(2A).$$ 
Based on Eqs.~(\ref{kf}) and (\ref{min}) it follows that $\epsilon^{\min}$ 
is independent on $\{\kappa,\delta\}$ and its value does not exceed 
$\frac{g\sqrt{2}}{4}(1+\gamma_{d}/\gamma)^{-1/2}$. 
The cavity output field, i.e., the number of photons escaping the cavity per second, can be evaluated via 
$\kappa \langle a^{\dagger}a\rangle_{s}$. Particularly, in Fig.~(\ref{fig2}) 
the minimum value of the steady-state mean-photon number is 
$\langle a^\dagger a\rangle^{\min}_{s} \approx 0.06$ and is 
achieved when $(\epsilon/\gamma)^{\min} \approx 0.54$. 
An explanation
of the steady-state behaviors shown in Fig.~(\ref{fig2}) can be found 
if one represents the mean-photon number given by (\ref{apm}) as follows:
\begin{eqnarray}
\langle a^{\dagger} a\rangle_{s}&=&\frac{\epsilon}{\kappa^{2}}
\{\epsilon + |g_{0}|\langle R_{z}\rangle_{s}\cos{\phi}\} + \frac{|g_{0}|}{\kappa(\kappa 
+ 2\Gamma_{+} + 2\Gamma_{-})} \nonumber \\
&\times&\{|g_{0}| + \epsilon\langle R_{z}\rangle_{s}\cos{\phi} -2(\Gamma_{+}-\Gamma_{-})
(|g_{0}|\langle R_{z}\rangle_{s} \nonumber \\
&+& \epsilon\cos{\phi})/\kappa\}, \label{cint}
\end{eqnarray}
where 
$$
\langle R_{z}\rangle_{s} =-(\Gamma_{+} - \Gamma_{-})/
(\Gamma_{+} + \Gamma_{-}).
$$ 
From the above expression (\ref{cint}), one can see that for $\delta=0$ 
the mean-photon number due to weaker external pumping of the cavity mode 
is proportional to $\epsilon^{2}$ while that due to stronger driving of 
the two-level qubit to $|g_{0}|^{2}$, respectively. 
There is also a cross-contribution proportional to $\epsilon |g_{0}|\cos{\phi}$. 
All these terms demonstrate interference effects among the contributions 
due to two pumping lasers and, hence, the minima's nature in Fig.~(\ref{fig2}).
In particular, for $\Gamma_{+} \gg \Gamma_{-}$ one has from Eq.~(\ref{cint}) that
\begin{eqnarray}
\langle a^{\dagger}a\rangle_{s} \approx \frac{\epsilon^{2}}{\kappa^{2}} + \frac{|g_{0}|^{2}}{\kappa^{2}} - \frac{2\epsilon|g_{0}|}{\kappa^{2}}\cos{\phi},
\label{cint1}
\end{eqnarray}
while for $\Gamma_{-} \gg \Gamma_{+}$ we have, respectively,
\begin{eqnarray}
\langle a^{\dagger}a\rangle_{s} \approx \frac{\epsilon^{2}}{\kappa^{2}} + \frac{|g_{0}|^{2}}{\kappa^{2}} + \frac{2\epsilon|g_{0}|}{\kappa^{2}}\cos{\phi}.
\label{cint2}
\end{eqnarray}
Thus, indeed, the output field suppression (or enhancement) occurs because of the interference effect taking place among the fraction of light 
$|g_{0}|^{2}/\kappa^{2}$ scattered by the atom into the cavity mode due to stronger pumping by the first laser beam and, respectively, 
the photon field of the second weaker laser field characterized by $\epsilon^{2}/\kappa^{2}$ (see, also, Figure~\ref{fig1}). Furthermore,
the nature of destructive or constructive interference can be understood as follows: In the dressed-state picture both lasers are simultaneously in 
resonance with the dressed-state transitions $|+\rangle \leftrightarrow |+\rangle$ and $|-\rangle \leftrightarrow |-\rangle$ and, hence, different 
signs in front of the last term in Eqs.~(\ref{cint1},\ref{cint2}). Actually, if $\Gamma_{+} \gg \Gamma_{-}$ the atom is located on the lower dressed-state 
$|-\rangle$ while it resides on the higher dressed-state $|+\rangle$ when $\Gamma_{-} \gg \Gamma_{+}$. However, the destructive interference can be 
turned into constructive one (or viceversa) via varying the phase difference $\phi$ (see Eq.~\ref{cint1} and Eq.~\ref{cint2}). This allows a better 
control of the output cavity field. Notice that on resonance, i.e., $\Delta=0$, the inhibition effects are absent 
when $\gamma_{+}=\gamma_{-}$ because $\Gamma_{+}=\Gamma_{-}$ resulting in $\langle R_{z}\rangle_{s}=0$ 
(see Eq.~\ref{cint}). Also, one can 
obtain small values for $\langle a^{\dagger}a\rangle_{s}$ if $\kappa > \gamma$. 
However, in this case we are in the bad-cavity limit and, therefore, 
lower values for mean-photon number or even zero are expected \cite{cbm,zb_sc,swn}. 
Thus, in contrary, the cavity output field suppression reported here occurs 
in the good-cavity limit, i.e., when $\gamma > \kappa$ and $g> \{\kappa,\gamma\}$.
\begin{figure}[t]
\includegraphics[width = 8cm]{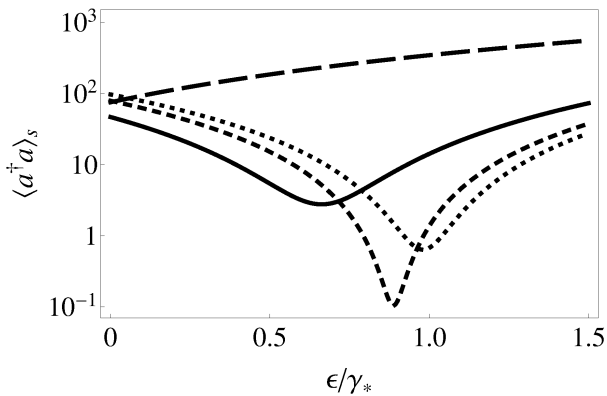}
\caption{\label{fig3}  
The steady-state dependences of the mean-photon number 
$\langle a^{\dagger}a\rangle_{s}$ as a function of $\epsilon/\gamma_{\ast}$.
The solid-line is for $\gamma_{\ast} \equiv \gamma_{+}=\gamma_{-}$, 
while the long-dashed curve stands for $\gamma_{\ast} = \gamma_{-}$ 
and $\gamma_{+} \to 0$. Further, the short-dashed line is for 
$\gamma_{\ast} = \gamma_{+}$ and $\gamma_{-} \to 0$, whereas the 
dotted curve corresponds to $\Delta=\delta=0$. Other parameters 
are the same as in Fig.~(\ref{fig2}) with $\Delta/\Omega=1$ and 
$\delta/\gamma_{\ast}=0$.}
\end{figure}

Apart from the dependence of the expression (\ref{cint}) on the parameters $\{\epsilon,\kappa,\phi, |g_{0}|\}$ it depends also on the generalized 
dressed-state decay rates $\Gamma_{\pm}$. These decay rates can be modified either by varying the detuning $\Delta$ in free-space with 
$\gamma_{+}=\gamma_{-}$ or via modification of the density of modes at the dressed-state frequencies $\omega \pm 2\bar \Omega$ and, consequently,  
$\gamma_{+} \not = \gamma_{-}$ which is a typical situation in photonic crystal environments  \cite{john,me_chk,gx}, for instance. In particular, one 
can have also the situation when $\gamma_{+} \gg \gamma_{-}$ or $\gamma_{-} \gg \gamma_{+}$. Figure (\ref{fig3}) shows the mean-photon numbers 
obtained with the help of the expression (\ref{cint}) when the two-level emitter is located inside a microscopic cavity engineered in a photonic crystal material. 
In this case, the output cavity field can be suppressed even on atom-laser frequency resonance, i.e. when $\Delta=0$ (see the dotted curve), because 
$\gamma_{+} \not = \gamma_{-}$ and the population will be distributed unequally among the dressed states while 
\begin{eqnarray*}
\langle R_{z}\rangle_{s}=\frac{\gamma_{-}-\gamma_{+}}{
\gamma_{-} + \gamma_{+} + 2\gamma_{d}} \not = 0, ~~~{\rm if ~~\Delta=0}.
\end{eqnarray*}
Negative values for the dressed-state inversion with $\phi=0$ lead to cavity output field suppression (see Fig.~\ref{fig3} and Eq.~\ref{cint1}). For the sake 
of comparison, the solid curve stands for ordinary vacuum-cavity environments. Thus, finalizing, we have shown here how the output cavity field can be 
minimized due to interference effects.

\section{Summary}
Summarizing, we have demonstrated the feasibility of cavity output field control via interference effects. The system of interest 
is formed from a strongly pumped two-level atom placed in an optical micro-resonator. A second weak laser being in resonance 
with the cavity mode frequency is probing the whole system through one of the cavity's mirror. Consequently, interference effects 
occur among the light scattered in the cavity mode by the strongly pumped atom and the incident weaker laser field leading to output 
cavity field inhibition or enhancement. Furthermore, the destructive interference can be turned into constructive one (or viceversa) 
via varying the phase difference of the applied lasers providing in this way a better control over output electromagnetic field.The idea 
works for photonic crystal environments as well.

\acknowledgments
We acknowledge the financial support by the German Federal Ministry 
of Education and Research, grant No. 01DK13015, and Academy of 
Sciences of Moldova, grant No. 13.820.05.07/GF. Furthermore we are 
grateful for the hospitality of the Theory Division of the 
Max-Planck-Institute for Nuclear Physics from Heidelberg, Germany.


\end{document}